\documentclass{elsart}

\usepackage{graphics}

\usepackage{amssymb,amsmath}

\newcommand{\conf}[2]{\ensuremath{(A, #1; B, #2)}}
\newcommand{\T}[1]{\ensuremath{\overline{#1}}}
\newcommand{\emptytuple}{\ensuremath{\langle\rangle}}
\newcommand{\semijoin}{\ensuremath{\ltimes}}
\newcommand{\ARITY}{\ensuremath{\mathrm{arity}}}
\newcommand{\game}[3]{\ensuremath{G_{#1}\conf{#2}{#3}}}

\begin{document}

\begin{frontmatter}


\title{On the expressive power of semijoin queries}
\author[LUC]{Dirk Leinders\corauthref{corr}}
\ead{dirk.leinders@luc.ac.be} \author[Warsaw]{Jerzy
  Tyszkiewicz\thanksref{jurek}} \ead{jty@mimuw.edu.pl}
\author[LUC]{Jan Van den Bussche} \ead{jan.vandenbussche@luc.ac.be}
\corauth[corr]{Corresponding author.}  \address[LUC]{Limburgs
  Universitair Centrum, Universitaire Campus, 3590 Diepenbeek,
  Belgium} \address[Warsaw]{Institute of Informatics, Warsaw
  University, ul.~Banacha 2, 02-097 Warszawa, Poland }
\thanks[jurek]{This author has been partially supported by the
  European Community Research Training Network ``Games and Automata
  for Synthesis and Validation'' (GAMES), contract
  HPRN-CT-2002-00283.}

\begin{abstract}
  The semijoin algebra is the variant of the relational algebra
  obtained by replacing the join operator by the semijoin operator. We
  provide an Ehrenfeucht-Fraiss\'{e} game, characterizing the
  discerning power of the semijoin algebra. This game gives a method
  for showing that queries are not expressible in the semijoin
  algebra.
\end{abstract}

\end{frontmatter}

\section{Introduction}
\label{sec:introduction}

Semijoins are very important in the field of database query
processing. While computing project-join queries in general is
NP-complete in the size of the query and the database, this can be
done in polynomial time when the database schema is
acyclic~\cite{Yannakakis81}, a property known to be equivalent to the
existence of a semijoin program~\cite{BFMY83}. Semijoins are often
used as part of a query pre-processing phase where dangling tuples are
eliminated.  Another interesting property is that the size of a
relation resulting from a semijoin is always linear in the size of the
input. Therefore, a query processor will try to use semijoins as often
as possible when generating a query plan for a given query (a
technique known as ``pushing projections''~\cite{GMUW00}). Also in
distributed query processing, semijoins have great importance, because
when a database is distributed across several sites, they can help
avoid the shipment of many unneeded tuples.

Because of its practical importance, we would like to have a clear
knowledge of the capabilities and the limitations of semijoins. For
example, Bernstein, Chiu and Goodman~\cite{BC81,BG81}
have characterized the conjunctive queries computable by semijoin
programs. In this paper, we consider the much larger class of queries
computable in the variant of the full relational algebra obtained by
replacing the join operator by the semijoin operator. We call this the
semijoin algebra (SA). We will define an Ehrenfeucht-Fraiss\'{e} game,
that characterizes the discerning power of the semijoin algebra. Using
this tool, we illustrate the borderline of expressibility of SA.

\section{Preliminaries}
\label{sec:preliminaries}

In this section, we give a formal definition of the semijoin algebra.

From the outset, we assume a universe $\mathbb{U}$ of basic data
values, over which a number of predicates or relations are defined.
These predicates can be combined into quantifier-free first-order
formulas, which are used in selection and semijoin conditions. The
names of these predicates and their arities are collected in the
vocabulary $\Omega$.  The equality predicate ($=$) is always in
$\Omega$. A database schema is a finite set $\mathbf{S}$ of relation
names, each associated with its arity.  $\mathbf{S}$ is disjoint from
$\Omega.$ A database $D$ over $\mathbf{S}$ is an assignment of a
finite relation $D(R) \subseteq \mathbb{U}^n$ to each $R \in
\mathbf{S}$, where $n$ is the arity of $R$.

\begin{defn}[Semijoin algebra, SA] Let $\mathbf{S}$ be a database
  schema. Syntax and semantics of the Semijoin Algebra is inductively
  defined as follows:
  \begin{enumerate}
  \item Each relation $R \in \mathbf{S}$ belongs to SA.
  \item If $E_1, E_2 \in $ SA have arity $n$, then also $E_1 \cup
    E_2$, $E_1 - E_2$ belong to SA and are of arity $n$.
  \item If $E_1\in $ SA has arity $n$ and $X$ is a subset of $ \{
    1,\ldots,n\}$, then $\pi_X(E_1)$ belongs to SA and is of arity
    $\#X$.
  \item If $E_1, E_2 \in$ SA have arities $n$ and $m$, respectively, and
    $\theta_1(x_1,\ldots,x_n)$ and
    $\theta_2(x_1,\ldots,x_n,y_1,\ldots,y_m)$ are quantifier-free
    formulas over $\Omega$, then also $\sigma_{\theta_1} (E_1)$ and
    $E_1 \semijoin_{\theta_2} E_2$ belong to SA and are of arity $n.$
  \end{enumerate}

  The semantics of the selection and the semijoin operator are as
  follows: $\sigma_{\theta_1} (E) := \{(a_1,\ldots,a_n)\in E \mid
  \theta_1(a_1,\ldots, a_n)$ holds$\}$, $E_1 \semijoin_{\theta_2}
  E_2 := \{ (a_1,\ldots,a_n) \in E_1 \mid \exists (b_1,\ldots, b_m)
  \in E_2$, $\theta_2(a_1,\ldots,a_n,b_1,\ldots,b_m)$ holds$\}$. The
  semantics of the other operators are well known.
\end{defn}

\section{An Ehrenfeucht-Fraiss\'{e} game for the semijoin algebra}
\label{sec:semijoingame}

In this section, we describe an Ehrenfeucht-Fra\"{\i}ss\'{e} game that
characterizes the discerning power of the semijoin algebra.

Let $A$ and $B$ be two databases over the same schema $\mathbf{S}$. The
\emph{semijoin game} on these databases is played by two players,
called the spoiler and the duplicator. They, in turn, choose tuples
from the tuple spaces $T_A$ and $T_B$, which are defined as follows:
$T_A := \bigcup_{R \in \mathbf{S}} \bigcup \big\{ \pi_{\mathrm{X}}(A(R))
  \mid \mathrm{X} \subseteq \{1,\ldots,\ARITY(R)\} \big\}$, and $T_B$
is defined analogously. So, the players can pick tuples from the
databases and projections of these.
 
At each stage in the game, there is a tuple $\T{a} \in T_A$ and a
tuple $\T{b} \in T_B$. We will denote such a configuration by
\conf{\T{a}}{\T{b}}.  The conditions for the duplicator to win the
game with 0 rounds are:

\begin{enumerate}
\item $\forall R \in \mathbf{S}, \forall \mathrm{X} \subseteq
  \{1,\ldots,\ARITY(R) \} : \T{a} \in \pi_{\mathrm{X}}(A(R))
  \Leftrightarrow \T{b} \in \pi_{\mathrm{X}}(B(R))$
\item for every atomic formula (equivalently, for every
  quantifier-free formula) $\theta$ over $\Omega$, $\theta(\T{a})$
  holds iff $\theta(\T{b})$ holds.
\end{enumerate} 

In the game with $m \geq 1$ rounds, the spoiler will be the first one
to make a move. Therefore, he first chooses a database ($A$ or $B$).
Then he picks a tuple in $T_A$ or in $T_B$ respectively. The
duplicator then has to make an ``analogous'' move in the other tuple
space.  When the duplicator can hold this for $m$ times, no matter
what moves the spoiler takes, we say that the duplicator wins the
$m$-round semijoin game on $A$ and $B$.  The ``analogous'' moves for
the duplicator are formally defined as legal answers in the next
definition.
\begin{defn}[legal answer]
  Suppose that at a certain moment in the semijoin game, the
  configuration is \conf{\T{a}}{\T{b}}. If the spoiler takes a tuple
  $\T{c} \in T_A$ in his next move, then the tuples $\T{d} \in T_B$,
  for which the following conditions hold, are legal answers for the
  duplicator:
\begin{enumerate}
\item $\forall R \in \mathbf{S}, \forall \mathrm{X} \subseteq
  \{1,\ldots,\ARITY(R) \}: \T{d} \in \pi_{\mathrm{X}}(B(R))
  \Leftrightarrow \T{c} \in \pi_{\mathrm{X}}(A(R))$
\item for every atomic formula $\theta$ over $\Omega$,
  $\theta(\T{a},\T{c})$ holds iff $\theta(\T{b},\T{d})$ holds.
\end{enumerate}
If the spoiler takes a tuple $\T{d} \in T_B$, the legal answers $\T{c}
\in T_A$ are defined identically.
\end{defn}

In the following, we denote the semijoin game with initial
configuration \conf{\T{a}}{\T{b}} and that consists of $m$ rounds, by
\game{m}{\T{a}}{\T{b}}. 

We first state and prove
\begin{prop} \label{lem:S_m}
  If the duplicator wins \game{m}{\T{a}}{\T{b}}, then for each
  semijoin expression $E$ with $\leq m$ nested semijoins and
  projections, we have $ \T{a} \in E(A) \Leftrightarrow \T{b} \in
  E(B) $.
\end{prop}
\begin{pf}
  We prove this by induction on $m$. The base case $m=0$ is clear. Now
  consider the case $m > 0$. Suppose that $\T{a} \in E_1
  \semijoin_{\theta} E_2$ but $\T{b} \not\in E_1 \semijoin_{\theta}
  E_2$. Then $\T{a} \in E_1(A)$ and $\exists \T{c} \in E_2(A):
  \theta(\T{a},\T{c})$, and either (*)~$\T{b} \not\in E_1(B)$ or
  (**)~$\lnot\exists \T{d} \in E_2(B): \theta(\T{b},\T{d})$. In
  situation (*), \T{a} and \T{b} are distinguished by an expression
  with $m-1$ semijoins or projections, so the spoiler has a winning
  strategy; in situation (**), the spoiler has a winning strategy by
  choosing this $\T{c} \in E_2(A)$ with $\theta(\T{a},\T{c})$,
  because each legal answer of the duplicator \T{d} has
  $\theta(\T{b},\T{d})$ and therefore $\T{d} \not\in E_2(B)$. So, the
  spoiler now has a winning strategy in the game
  \game{m-1}{\T{c}}{\T{d}}. In case a projection distinguishes \T{a}
  and \T{b}, a similar winning strategy for the spoiler exists.  In
  case \T{a} and \T{b} are distinguished by an expression that is
  neither a semijoin, nor a projection, there is a simpler expression
  that distinguishes them, so the result follows by structural
  induction.
\end{pf}

We now come to the main theorem of the text. This theorem concerns the
game \game{\infty}{\T{a}}{\T{b}}, which we also abbreviate as
\game{}{\T{a}}{\T{b}}. We say that the duplicator wins
\game{}{\T{a}}{\T{b}} if the spoiler has no winning strategy. This
means that the duplicator can keep on playing forever, choosing legal
answers for every move of the spoiler. 

\begin{thm}
  The duplicator wins \game{}{\T{a}}{\T{b}} if and only if for each
  semijoin expression $E$, we have $\T{a} \in E(A) \Leftrightarrow
  \T{b} \in E(B) $.
\end{thm}
\begin{pf}
  The `only if' direction of the proof follows directly from
  Proposition~\ref{lem:S_m}, because if the duplicator wins
  \game{}{\T{a}}{\T{b}}, he wins \game{m}{\T{a}}{\T{b}} for every $m
  \geq 0$. So, \T{a} and \T{b} are indistinguishable through all
  semijoin expressions. For the `if' direction, it is sufficient to
  prove that if the duplicator loses, \T{a} and \T{b} are
  distinguishable. We therefore construct, by induction, a semijoin
  expression $E_{\T{a}}^r$ such that (i) $\T{a} \in E_{\T{a}}^r(A) $,
  and (ii) $\T{b} \in E_{\T{a}}^r(B)$ iff the duplicator wins
  \game{r}{\T{a}}{\T{b}}. We define $E_{\T{a}}^0$ as
\[
\sigma_{\theta_{\T{a}}} \big( \bigcap_{R \in \mathbf{S}} \bigcap_{
  \{\mathrm{X} \subseteq Z \mid \T{a} \in \pi_{\mathrm{X}}(A(R)) \}}
\pi_{\mathrm{X}}(R) \big) - \bigcup_{R \in \mathbf{S}} \bigcup_{ \{
  \mathrm{X} \subseteq Z \mid \T{a} \not\in \pi_{\mathrm{X}}(A(R)) \}
} \pi_{\mathrm{X}}(R)
\]
In this expression, $Z$ is a shorthand for $\{ 1,\ldots, \ARITY(R)\}$
and $\theta_{\T{a}}$ is the \emph{atomic type} of \T{a} over $\Omega$,
i.e., the conjunction of all atomic and negated atomic formulas over
$\Omega$ that are true of \T{a}. 
 
We now construct $E_{\T{a}}^r$ in terms of $E_{\T{a}}^{r-1}$:
\[
\bigcap_{\T{c} \in T_A} \big( E_{\T{a}}^0
\semijoin_{\theta_{\T{a},\T{c}}} E_{\T{c}}^{r-1} \big) \cap \big(
E_{\T{a}}^0 - \bigcup_{j=1}^{s} \bigcup_{\theta} \big( E_{\T{a}}^0
\semijoin_{\theta} \bigcap_{\substack{\T{c} \in T_A\\
    \theta(\T{a},\T{c})}} (E_{\T{c}}^{r-1})^{\mathrm{compl}} \big) \big)
\]
In this expression, $\theta_{\T{a},\T{c}}$ is the atomic type of \T{a}
and \T{c} over $\Omega$; $s$ is the maximal arity of a relation in
$\mathbf{S}$; $\theta$ ranges over all atomic $\Omega$-types of two
tuples, one with the arity of \T{a}, and one with arity $j$. The
notation $E^{\mathrm{compl}}$, for an expression of arity $k$, is a
shorthand for
\[
E - \bigcup_{R \in \mathbf{S}} \bigcup_{ \substack{X \subseteq
    \{1,\ldots,\ARITY(R)\} \\ \#X = k}} \pi_{\mathrm{X}} (R)
\]
\end{pf}

\section{The expressive power of the semijoin algebra}
\label{sec:delineate}

In this section, we present some queries that delineate the expressive
power of the semijoin algebra. They are summarized in
Table~\ref{tab:delineate}. The operation $R \circ S$ for binary
relations $R$ and $S$ is a shorthand for $\pi_{1,4}
\big(\sigma_{2=3}(R \times S)\big)$.

\begin{table}
  \centering
  \caption{Queries delineating the expressive power of the semijoin algebra.}
  \begin{tabular}{|l|l|}
    \hline
    Expressible & Inexpressible\\
    \hline
    \hline
    $R \times S\ \cap\ T$     & $R \times S \subseteq T$\\
    $T \subseteq R \times S$  & $T = R \times S$\\
    & $R \circ S\ \cap\ T$\\
    & $T \subseteq R \circ S$\\
    & $R \circ S \subseteq T$\\
    \hline
    $\exists$ path of length $k$ & \\
    $\exists$ simple path of length $k$ ($k \leq 2$) & $\exists$
    simple path of length $k$ ($k \geq 3$)\\
    \hline
    $\exists$ cycle of length $k$ ($k \leq 2$) & $\exists$ cycle of length $k$ ($k \geq 3$)\\
    \hline
          & $\exists \geq k$
            elements 
            ($k \geq 3$)\\
    \hline
  \end{tabular}
  \label{tab:delineate}
\end{table}

We now discuss the results presented in the table. The semijoin
algebra lacks the cartesian product operator, but nevertheless one can
check if $T \subseteq R \times S$. Indeed, $T \subseteq R \times S$
iff $T - ( T\ \cap\ R \times S) = \emptyset$, and $T\ \cap\ R \times S
= (T \ltimes_{x_1=y_1\land x_2=y_2} R)
\ltimes_{x_3=y_1\land x_4=y_2} S$. Conversely, it is impossible
to check if $T \supseteq R \times S$. In
Figure~\ref{fig:TsubsumesRxS}, two databases $A$ and $B$ are shown
that are indistinguishable through semijoin expressions because the
duplicator has an obvious winning strategy. But $A$ satisfies $T
\supseteq R \times S$ and $B$ does not. The same databases actually
show that it is impossible to check if $T = R \times S$.

\begin{figure}
  \centering
  \begin{tabular}{cc}
    \begin{tabular}{cc}
      \begin{tabular}{c}
        \begin{tabular}{|c|}
          \hline
          $A(R)$\\
          \hline
          a\\
          b\\
          \hline
        \end{tabular}\\
        \\
        \begin{tabular}{|c|}
          \hline
          $A(S)$\\
          \hline
          1\\
          2\\
          \hline
        \end{tabular}
      \end{tabular}
      & 
      \begin{tabular}{|c|c|}
        \hline
        \multicolumn{2}{|c|}{$A(T)$}\\
        \hline
        a & 1\\
        a & 2\\
        b & 1\\
        b & 2\\
        \hline
      \end{tabular}
    \end{tabular}
    &
\hspace{1.0cm}
    \begin{tabular}{cc}
      \begin{tabular}{c}
        \begin{tabular}{|c|}
          \hline
          $B(R)$\\
          \hline
          a\\
          b\\
          c\\
          \hline
        \end{tabular}\\
        \\
        \begin{tabular}{|c|}
          \hline
          $B(S)$\\
          \hline
          1\\
          2\\
          3\\
          \hline
        \end{tabular}
      \end{tabular}
      &
      \begin{tabular}{|c|c|}
        \hline
        \multicolumn{2}{|c|}{$B(T)$}\\
        \hline
        a & 1\\
        a & 2\\
        b & 2\\
        b & 3\\
        c & 1\\
        c & 3\\
        \hline
      \end{tabular} 
    \end{tabular}
  \end{tabular}
  \caption{In $A$, $T = R \times S$, but not in $B$.}
  \label{fig:TsubsumesRxS}
\end{figure}

Although one can check in SA if a relation is contained in a
\textit{cartesian product}, it is impossible to check if a relation is
contained in or subsumed by a \textit{join}. Using our semijoin game,
one can show that databases $A$ and $B$ in
Figure~\ref{fig:TcontsubRcircS} satisfy the same semijoin expressions.
But $A$ satisfies $T = R \circ S$, while $B$ satisfies neither $T
\subseteq R \circ S$ nor $T \supseteq R \circ S$. Note that a binary
relation $R$ is transitive if and only if $R \circ R \subseteq R$.
This is a special case of $R \circ S \subseteq T$; yet, a similar
argument shows that transitivity is also inexpressible in the semijoin
algebra.

\begin{figure}
  \centering
  \begin{tabular}{cc}
    \begin{tabular}{ccc}
      \begin{tabular}{|c|c|}
        \hline
        \multicolumn{2}{|c|}{$A(R)$}\\
        \hline
        1 & a\\
        3 & b\\
        \hline
      \end{tabular}
      &
      \begin{tabular}{|c|c|}
        \hline
        \multicolumn{2}{|c|}{$A(S)$}\\
        \hline
        a & 2\\
        b & 4\\
        \hline
      \end{tabular}
      &
      \begin{tabular}{|c|c|}
        \hline
        \multicolumn{2}{|c|}{$A(T)$}\\
        \hline
        1 & 2\\
        3 & 4\\
        \hline
      \end{tabular}
    \end{tabular}
    &
\hspace{0.5cm}
    \begin{tabular}{ccc}
      \begin{tabular}{|c|c|}
        \hline
        \multicolumn{2}{|c|}{$B(R)$}\\
        \hline
        1 & a\\
        3 & b\\
        \hline
      \end{tabular}
      &
      \begin{tabular}{|c|c|}
        \hline
        \multicolumn{2}{|c|}{$B(S)$}\\
        \hline
        b & 2\\
        a & 4\\
        \hline
      \end{tabular}
      &
      \begin{tabular}{|c|c|}
        \hline
        \multicolumn{2}{|c|}{$B(T)$}\\
        \hline
        1 & 2\\
        3 & 4\\
        \hline
      \end{tabular}
    \end{tabular}
  \end{tabular}
  \caption{In $A$, $T = R \circ S$, but in $B$ neither $T \subseteq R
    \circ S$ nor $T \supseteq R \circ S$.}
  \label{fig:TcontsubRcircS}
\end{figure}

The existence of a path of length $k$ can be checked with the
following inductively defined semijoin expression:
\[ \left\{ \begin{array}{rcl} 
            \mathrm{path}(1) & := & R\\
            \mathrm{path}(k) & := & R \ltimes_{x_2=y_1} 
                                    \big(\mathrm{path}(k-1)\big)
          \end{array} 
        \right.
\]
Problems arise when we require the path to be simple. Let $D^{(k)}$ be
the structure $\{(1,2),(2,3),\ldots,(k-1,k),(k,1) \}$ over the schema
$\mathbf{S}$ containing a single edge relation $R$. Then, the
duplicator has a winning strategy in the infinite game played on
$D^{(k)}$ and $D^{(k+1)}$ where $k \geq 4$. To see this, note that
only three types of moves are possible here: next tuple (change only
first component of pebbled tuple), previous tuple (change only second
component) and other tuple (change both components). The duplicator
can answer every type of move of the spoiler. But $D^{(k+1)}$ contains
a simple path of length $k$ and $D^{(k)}$ does not. For $k=3$, note
that $D^{(3)}$ and $D^{(4)}$ \textit{are} distinguishable.
Nevertheless, existence of a simple path of length 3 is still
inexpressible because $D^{(4)}$ is indistinguishable from the
structure consisting of two disjoint copies of $D^{(3)}$. For $k=2$,
the existence of a path of length 2 is expressible as $R
\ltimes_{x_2=y_1\land x_2\neq x_1\land y_2\neq x_2} R$.

Another property that is inexpressible in SA is the existence of a
cycle of length $k$. For $k \geq 4$, the inexpressibility result
follows because $D^{(k)}$ contains a cycle of length $k$ and
$D^{(k+1)}$ does not. For $k=3$, that the structure consisting of two
disjoint copies of $D^{(3)}$ contains a cycle of length 3, but
$D^{(4)}$ does not.

A last example of a query that is inexpressible in SA is the query
that asks if there are at least $k$ elements in a unary relation $S$,
where $k \geq 3$. This property is inexpressible because the
duplicator has a winning strategy in the infinite game played on two
relations, one with 2 and one with $k$ distinct elements.

\section{Impact of order}
\label{sec:impact-order}

In this section, we investigate the impact of order. On ordered
databases (where $\Omega$ now also contains a total order on the
domain), the query that asks if there are at least $k$ elements in a
unary relation $S$ becomes expressible as $\mathrm{at\_least}(k)$,
which is inductively defined as follows:
\[ \left\{ \begin{array}{rcl}
    \mathrm{at\_least}(1) & := & S\\
    \mathrm{at\_least}(k) & := & S \ltimes_{x_1<y_1}
    \big(\mathrm{at\_least}(k-1)\big)
          \end{array} 
        \right.
\]
Note that this query is independent of the order. This is very
interesting because in first-order logic, there also exists an
order-invariant query that is expressible with but inexpressible
without order (\cite[Exercise 17.27]{AHV95} and \cite[Proposition
2.5.6]{EF99}).

Some inexpressible queries presented in Section~\ref{sec:delineate}
remain inexpressible on ordered databases. An example is the query $R
\times S \subseteq T$. Indeed, consider the following databases $A$
and $B$: $A(R)=B(R)= \{1,2,\ldots,m \}$, $A(S)=B(S)= \{
m+1,m+2,\ldots, 2m \}$, $A(T)=A(R)\times A(S)$ and $B(T)=A(T) -
\{(\frac{m+1}{2}, m+\frac{m+1}{2}) \}$. We will show that when
$m=2n+1$, the duplicator has a winning strategy in the $n$-round
semijoin game \game{n}{\emptytuple}{\emptytuple} with $\Omega =
\{=,<\}$. From Lemma~\ref{lem:S_m}, it then follows that the query $R
\times S \subseteq T$ is inexpressible in SA. The duplicator's winning
strategy consists of playing exact match until the spoiler chooses
\T{c} to be the special tuple $(\frac{m+1}{2}, m+\frac{m+1}{2})$ in
$A$. In that case we must distinguish five possibilities for the
previous tuple \T{a}: (1) $a_1 = \frac{m+3}{2}$, (2) $a_1 =
\frac{m-1}{2}$, (3) $a_1 = \frac{m+1}{2}$ and $a_2 = m+\frac{m+3}{2}$,
(4) $a_1 = \frac{m+1}{2}$ and $a_2 = m+\frac{m-1}{2}$ and (5) all
other cases. The duplicator chooses \T{d} equal to
$(\frac{m-1}{2},m+\frac{m+1}{2})$ in case 1, $(\frac{m+3}{2},
m+\frac{m+1}{2})$ in case 2, $(\frac{m+1}{2},m+\frac{m-1}{2})$ in case
3, $(\frac{m+1}{2},m+\frac{m+3}{2})$ in case 4, and
$(\frac{m-1}{2},m+\frac{m+1}{2})$ in case 5. Let us assume case 1
applies; cases 2 to 5 are analogous. Then, there are two
possibilities. First, if the spoiler chooses a value $c_1 \neq a_1 -
1$ or if he chooses a value $d_1 \neq b_1 + 1$ in some next round, the
duplicator can play exact match and the game starts over. Second, if
the spoiler chooses in each next round $c_1 = a_1 - 1$ or $d_1 = b_1 +
1$, the duplicator answers $d_1 = b_1 - 1$ or $c_1 = a_1 + 1$,
respectively. The duplicator can follow this strategy for at least
$\frac{m-3}{2}=n-1$ rounds. Counting from the round where the spoiler
chose the special tuple, we thus see that the duplicator wins the game
\game{n}{\emptytuple}{\emptytuple}.

Exactly the same argument shows that also the query $R \times S = T$
is inexpressible in SA with order.

Another query from Table~\ref{tab:delineate} that remains
inexpressible in SA with order is $R \circ S \subseteq T$. Therefore,
consider the following databases $A$ and $B$: $A(R) = B(R) =
\{1,\ldots,m\} \times \{2m+1\}$, $A(S) = B(S) = \{2m+1\} \times
\{m+1,\ldots,2m\}$, $A(T) = A(R) \circ A(S) = \{1,\ldots,m \} \times
\{m+1,\ldots,2m \}$ and $B(T)= B(R) \circ B(S)\ -\ \{(\frac{m+1}{2},
m+\frac{m+1}{2}) \}$. A similar argument as in the previous paragraph
shows that when $m=2n+1$, the duplicator wins
\game{n}{\emptytuple}{\emptytuple}. Again, this also shows that $R
\circ S = T$ is inexpressible in SA with order.

For the remaining SA-inexpressible queries in
Table~\ref{tab:delineate}, the question whether they become
expressible in SA with order remains open.

\section{Concluding remarks}
\label{sec:concluding-remarks}

Interestingly, there is a fragment of first-order logic very similar
to the semijoin algebra: it is the so called ``guarded
fragment''(GF)~\cite{AvBN98}, which has been studied in the field
of modal logic. This is interesting because the motivations to study
this fragment came purely from the field of logic and had nothing to
do with database query processing.  Indeed, the purpose was to extend
propositional modal logic to the predicate level, retaining the good
properties of modal logic, such as the finite model property. An
important tool in the study of the expressive power of the GF is the
notion of ``guarded bisimulation'', which provides a characterization
of the discerning power of the GF.

When we only allow conjunctions of equalities to be used in the
semijoin conditions, SA is subsumed by GF, and conversely, every GF
sentence is expressible in SA.

When negations of equalities are allowed in semijoin conditions,
however, SA is no longer subsumed by GF. A counterexample is the query
that asks whether there are at least two distinct elements in a single
unary relation $S$. This is expressible in SA as $S \semijoin_{x_1
  \neq y_1} S$, but it is not expressible in GF. Proofs of the claims
presented in this section will be presented in a separate paper.

\bibliographystyle{plain} \bibliography{include/bibliography}

\end{document}